# Optimized routines for event generators in QED-PIC codes

V Volokitin[1,2], S Bastrakov[5], A Bashinov[3], E Efimenko[3], A Muraviev[3],
A Gonoskov[4,3,2] and I Meyerov[1,2]

[1] Mathematical Center, Lobachevsky University, Nizhny Novgorod, Russia
[2] Department of Software and Supercomputing Technologies, Lobachevsky University, Nizhny Novgorod, Russia
[3] Institute of Applied Physics, Russian Academy of Sciences, Russia
[4] Department of Physics, University of Gothenburg, Gothenburg, Sweden
[5] Helmholtz-Zentrum Dresden-Rossendorf, Dresden, Germany

Email: meerov@vmk.unn.ru

**Abstract.** In recent years, the prospects of performing fundamental and applied studies at the next-generation high-intensity laser facilities have greatly stimulated the interest in performing large-scale simulations of laser interaction with matter with the account for quantum electrodynamics (QED) processes such as emission of high energy photons and decay of such photons into electron-positron pairs. These processes can be modelled via probabilistic routines that include frequent computation of synchrotron functions and can constitute significant computational demands within accordingly extended Particle-in-Cell (QED-PIC) algorithms. In this regard, the optimization of these routines is of great interest. In this paper, we propose and describe two modifications. First, we derive a more accurate upper-bound estimate for the rate of QED events and use it to arrange local sub-stepping of the global time step in a significantly more efficient way than done previously. Second, we present a new high-performance implementation of synchrotron functions. Our optimizations made it possible to speed up the computations by a factor of up to 13.7 depending on the problem. Our implementation is integrated into the PICADOR and Hi-Chi codes, the latter of which is distributed publicly (https://github.com/hi-chi/pyHiChi).

## 1. Introduction

The numerical simulation of laser plasma with the Particle-in-Cell (PIC) method is an area of immediate interest in computational physics. Some of the pressing problems are rather computationally demanding and require large-scale simulation on supercomputers. An important computational property of the PIC method is the spatial locality of interactions, which provides possibilities for massively parallel computing. The last decades saw a great progress in the development of PIC codes for plasma simulation. It includes the improvement of performance and scalability within many codes, such as VLPL [1], OSIRIS [2], PIConGPU [3], Smilei [4], PICADOR [5, 6], EPOCH [7], WarpX [8] and many others. Such codes have demonstrated good performance and scaling on a variety of architectures, including CPUs, GPUs, Xeon Phi [9-11].

However, certain types of problems are still challenging for large-scale simulation due to distinctive algorithmic and numerical features. These include extended PIC simulations with account for the effects of strong field quantum electrodynamics (QED), the so-called QED-PIC simulations [12-18]. This kind of simulations is necessary for planning experiments at that next generation laser facilities that are currently under construction. These facilities are expected to achieve record values of electron-positron plasma density and electromagnetic energy density, as well as maximal energy of electrons and photons. This may open the door to exploring fundamental properties of matter and vacuum. Planning such experiments requires, among other contributions, the development of computational tools and their use for performing large-scale simulations.

The most common approach for such simulations is extending the PIC method with consideration of high-energy photon emission by electrons and positrons and the decay of such photons into electron-positron pairs, resulting in the so-called QED-PIC simulation model [19-21]. These processes are modeled using probabilistic procedures, with expressions for probabilities being derived from quantum electrodynamics [25]. Papers [21, 26] detail QED-PIC schemes with explicit formulas for the probabilities of these processes in terms of *synchrotron functions*. These schemes impose restrictions on the time step, depending on the intensity of the electromagnetic field. To significantly reduce the required computational resources, an automatic *local sub-stepping in time* is used. Thus, in regions of high field intensity, one time step of the PIC method can be locally subdivided making it possible to simulate multiple QED events within the single global time step. As customary for PIC simulations, the time step should sufficiently resolve the macroscopic (classical) electromagnetic field dynamics, and the fields and currents are updated once per time step [21].

Such 3D QED-PIC simulations pose a few computational challenges. First, QED effects often result in a significant portion of particles being localized in a very small part of the simulation area [14, 17, 18, 27, 30]. Second, the QED part of the simulation is an additional simulation stage, with very uneven computational costs depending on the local number of particles and electromagnetic field intensity. Our earlier work [21] describes the organization of the QED module implemented in the PICADOR and ELMIS codes, and paper [22] demonstrates the use of load balancing schemes to efficiently utilize the resources of multicore systems in such simulations. In this paper, we propose and describe two modifications to further boost the efficiency of QED-PIC simulations.

1. **Refinement of estimates used for optimization**. The characteristic time intervals of the outlined QED events can vary significantly and thus it is beneficial to use estimates for the rates of these events to avoid excessive calculations of the exact rates or, on the contrary, to ensure they are not disregarded when they are necessary. This has been implemented in the previous edition of our numerical routines. However, we have since discovered that the estimates used were overcautious and it is possible to significantly increase the time step in many cases. We derive more accurate estimates, which alone yielded 8 times speedup for the problem in question.
2. **High-performance implementation of synchrotron functions**. Our analysis showed that in QED-PIC simulations, the calculation of synchrotron functions using the GSL library takes more than 20% of the total computation time. In this regard, we developed a new high-performance implementation of synchrotron functions, which made it possible to speed up their calculation by a factor of 1.5 while maintaining a $10^{-14}$ accuracy, which is close enough to the accuracy of double-precision computations. We also showed that the approximation with an accuracy of $10^{-8}$ led to further performance improvement while yielding almost identical simulation results. The refinement of estimates in combination with the improved implementation of synchrotron functions yield a speedup of up to 13.7, which is substantial for QED-PIC simulations.

The paper is organized as follows. In Section 2 we give a short overview of the QED-PIC method. In Section 3 we present the two optimizations we have developed and discuss numerical results. Section 4 concludes the paper.

## 2. Methods

*2.1. Basic computational loop of the PIC method for laser-plasma simulation*
The PIC method [23, 24] simulates self-consistent dynamics of particles driven by electromagnetic fields and at the same time affecting those fields. When applied to laser plasma simulations, the method operates on two main data sets: the electromagnetic field defined on a Cartesian spatial grid (usually the Yee grid) and an ensemble of so-called macroparticles (being used to mitigate the computational costs needed for typically too large number of real particles). The basic computational loop has the following stages:

1. Solving Maxwell's equations to propagate values of the electric and magnetic fields on the spatial grid.
2. Field interpolation from nearest grid points to each particle position to compute the Lorenz force.
3. Computing momentums and positions of each particle for the next time step according to the Lorenz force (particle push).
4. Deposition of currents induced by particle movement to the spatial grid.

*2.2. Overview of the QED-PIC method*

As in the standard PIC method, a plasma is described using an ensemble of macroparticles, each representing a certain number of real particles of the same type. The electromagnetic radiation is described by simultaneously using two approaches: the classical representation via grid values of the field and the quantum representation using an ensemble of photons treated as macroparticles. Such a combination is needed due to the impossibility of providing sufficient grid resolution for high-frequency radiation of charged particles in the case of strong electromagnetic fields. Unlike the classical coherent radiation of plasma particles, individual incoherent quanta can be described as particles. The correctness of such a dual description for many problems and, in particular, the absence of double accounting of electromagnetic radiation, is shown in [21].

The processes of photon emission by electrons and positrons, as well as the decay of photons into electron-positron pairs, are probabilistic. Within a few approximations, such probabilities can be calculated analytically by means of quantum electrodynamics [25]. The probability depends on the electromagnetic field intensity, and the energy of the original photon and resulting particles. Thus, during the simulation these probabilities are calculated and the events of pair / photon generation are sampled accordingly. Apart from the most natural approach based on the inverse sampling method [20], there are the so-called alternative [19] and modified [21] event generators. The advantage of the modified event generator is that it correctly treats the entire radiation spectrum, resolving the issue of singularity of the energy distribution function of emitted photons in the low-energy limit. For this reason, we implemented this approach in the PICADOR and Hi-Chi codes used in this work.

The resulting QED-PIC computational loop has the same stages as the basic Particle-in-Cell loop, but with a modification of the particle push stage. At this stage, for each charged macroparticle (macroelectron, macropositron) we generate a new macrophoton with probability density $P(\delta)$:

$$P(\delta) = \left[\Delta t * \frac{e^2 mc}{\hbar^2}\right] * \frac{\sqrt{3}}{2\pi} \frac{\chi}{\gamma} \frac{1-\delta}{\delta} \left(F(z_q) + \frac{3}{2}\delta\chi z_q G(z_q)\right), \quad (1)$$

where $m$ is the electron mass, $e$ is the electron change, $\hbar$ is the reduced Planck constant, $c$ is the speed of light, $\Delta t$ is the time step, $\gamma$ is the relativistic Lorentz-factor of the particle, $\delta = \frac{\hbar\omega}{mc^2\gamma}$ is the ratio of photon energy to the full energy of the original particle $\varepsilon = mc^2\gamma$, $F(x)$ and $G(x)$ are the first and second synchrotron functions, $z_q = \frac{2}{3}\chi^{-1}\frac{\delta}{1-\delta}$, and $\chi \equiv \frac{e\hbar}{m^3c^4}\sqrt{\left(\frac{\varepsilon\vec{E}}{c} + \vec{p} \times \vec{H}\right)^2 - (\vec{p} \cdot \vec{E})^2}$ is a dimensionless parameter characterizing the transverse acceleration of the particle in the field. For electrons and positrons this parameter can be calculated as $\chi = \gamma \frac{H_{eff}}{E_s}$, where $E_s = \frac{m^2c^3}{e\hbar}$ is the Schwinger field and $H_{eff}$ is the effective field that acts on the particle. The generated photon is assumed to have the same direction of propagation as the parent particle.

For each macrophoton we sample events of decay into an electron-positron pair with probability $P_p(\delta_e)$. If that event occurs, we generate a new macroelectron and macropositron and remove the original macrophoton. The decay probability is calculated as

$$P_p(\delta_e) = \left[\Delta t * \frac{e^2 mc}{\hbar^2}\right] * \frac{\sqrt{3}}{2\pi} \frac{\chi_\gamma mc^2}{\hbar\omega} (1-\delta_e)\delta_e \left(F(z_p) - \frac{3}{2}\chi_\gamma z_p G(z_p)\right), \quad (2)$$

with $z_p = \frac{2}{3}\frac{1}{\chi_\gamma(1-\delta_e)\delta_e}$, $\delta_e = \frac{mc^2\gamma_e}{\hbar\omega}$ is the ratio of the electron's energy to the energy of the original photon, and $\chi_\gamma = \frac{\hbar\omega}{mc^2}\frac{H^\gamma_{eff}}{E_s}$. The direction of propagation of both generated particles is assumed to match that of the parent photon.

In strong fields the typical interval between such events can be much lower than the time step of the PIC simulation, which is normally chosen just to sufficiently resolve plasma and laser wavelength. Thus, we employ automatic local sub-stepping that only takes place on this stage of the simulation (in some areas), and does not affect the main time step. This adaptive event generator [21] allows simulating a cascade of events during a single time step. Algorithms 1 and 2 present the generator for particles and photons, respectively. A detailed description is given in [21].

---
**Algorithm 1.** Particle push stage with adaptive event generator for charged macroparticles, propagates a macroparticle by a single time step of the main PIC loop $\Delta t$.
1. **Calculate** time sub-step $sub_t$ to be used on the current time iteration for the current cell;
2. **For** $(i = 0; i < \frac{\Delta t}{sub_t}; i++)$
3.     **Propagate** the particle by $sub_t$ in time with the Boris pusher;
4.     **Compute** $H_{eff}$ and $\chi$;
5.     **Sample** $x, y \in U[0,1]$;
6.     **If** $(y < 3x^2 P(x^3))$
7.         **Emit** a macrophoton;

---
**Algorithm 2.** Particle push stage with adaptive event generator for macrophotons, propagates a macrophoton by a single time step of the main PIC loop $\Delta t$.
1. **Calculate** time sub-step $sub_t$ to be used on the current time iteration for the current cell;
2. **For** $(i = 0; i < \frac{\Delta t}{sub_t}; i++)$
3.     **Propagate** the photon by $sub_t$ in time with the speed of light;
4.     **Compute** $H^\gamma_{eff}$ and $\chi_\gamma$;
5.     **Sample** $x, y \in U[0,1]$;
6.     **If** $(y < P_p(x))$
7.         **Create** an electron-positron pair;
8.         **Delete** the current macrophoton and **stop** the algorithm;

---

### 3. Modifications of the QED routines

In certain simulations of interest [18, 27, 30], accounting for QED effects takes up to 95% of the total QED-PIC run time. The main hotspots are related to calling the particle pusher for each sub-step, and computing probabilities $P(\delta)$ and $P_p(\delta_e)$, which in turn mostly consists of computing the synchrotron functions. This paper addresses these issues.

*3.1. Refinement of estimates used for sub-stepping*

*3.1.1. Method*

This section presents our approach to reducing the number of sub-steps, and consequently the number of loop iterations in Algorithms 1 and 2. With a careful analysis of the underlying probability estimates it can be possible without a substantial loss of accuracy. As we show below, the estimates can be overcautious, which leads to over-refined sub-stepping, and thus offers a potential for optimization.

The sub-step value $\Delta t/sub_t$ in Algorithms 1 and 2 depends on two quantities [21]:
- The maximum decay probability $P_{\max}$.
- The upper bound estimate for the decay probability expression.

At first, we analyzed the actual frequency of QED events depending on the number of time sub-steps since the previous event presented at Figure 1, left. On average, 12% of the decay events happened after 10 sub-steps since the previous event.

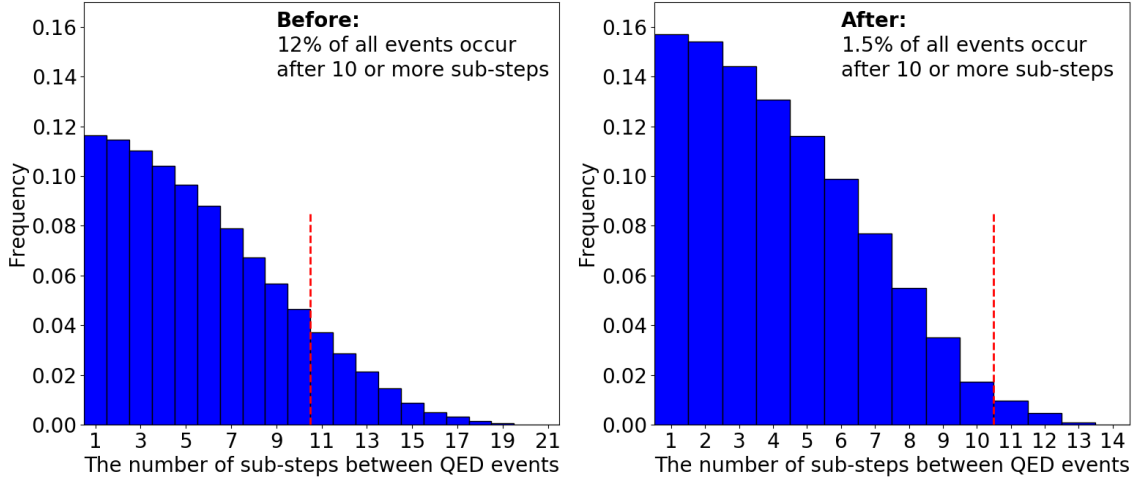

Figure 1. Histogram of QED events frequency depending on the number of time sub-steps since the previous event. Left: original implementation. Right: after refinement of the estimates.

This opens up a way to improve the overly conservative probability estimates. Consider the photon emission probability equation (1). According to Algorithm 1 we sample two uniformly distributed numbers $x, y \sim U[0,1]$ and compute $P_m(x) = 3x^2 P(x^3)$, then in the case $y < P_m(x)$ we emit a photon with energy $y$.

Let us now consider another random variable $z = \frac{P_m(x)}{y}$. The photon emission condition in terms of $z$ takes the form $P(1 < z) = P_{\max}$. With some natural assumptions that hold in real applications, we get the following:

$$P(1 < z) = \int_0^1 \int_0^{cx^2\left(\int_{\frac{x^3}{b(1-x^3)}}^{\infty} K_{\frac{5}{3}}(t)dt + \frac{x^6}{1-x^3}K_{\frac{2}{3}}\left(\frac{x^3}{b(1-x^3)}\right)\right)} dy\, dx =$$

$$= \int_0^1 cx^2 \left(\int_{\frac{x^3}{b(1-x^3)}}^{\infty} K_{\frac{5}{3}}(t)dt + \frac{x^6}{1-x^3}K_{\frac{2}{3}}\left(\frac{x^3}{b(1-x^3)}\right)\right) dx =$$

$$= \int_0^1 cx^2 \int_{\frac{x^3}{b(1-x^3)}}^{\infty} K_{\frac{5}{3}}(t)dt\, dx + \int_0^1 cx^2 \frac{x^6}{1-x^3} K_{\frac{2}{3}}\left(\frac{x^3}{b(1-x^3)}\right) dx,$$

where $b = \frac{3}{2}\chi$, $c = \frac{\sqrt{3}}{\pi\gamma}\left[\Delta t \frac{e^2 mc}{\hbar^2}\right]$, and $K_k(t)$ is the modified Bessel function of the second kind.

For these integrals, we use upper bound estimates of the Bessel function for different ranges of the parameter $b$, which yields the following overall upper bound estimate for the decay probability:

$$P(1 < z) < f(b) = \begin{cases} 0.56\pi cb + 0.1cb, & for\ b < 0.1 \\ 0.45\pi cb^{\frac{11}{12}} + 0.1cb, & for\ 0.1 < b < 0.5 \\ 0.42\pi cb^{\frac{19}{24}} + 0.1cb, & for\ 0.5 < b < 10 \\ 0.55\pi cb^{\frac{2}{3}} + 0.1cb, & for\ b > 10 \end{cases}$$

We applied this technique for $P_p(\delta_e)$ used in Algorithm 2, and found out that the same estimate for all values of the parameter $b$ can be used:

$$P(1 < z_p) < f(b) = 0.1cb$$

The histogram of the number of sub-steps between QED events after these modifications is shown at Figure 1, right. Due to the new estimates, only 1.5% of all events occur after 10 or more sub-steps, which shows our goal to reduce the number of loop iterations was achieved.

*3.1.2. Numerical results*

To verify the optimized probability estimates, we performed QED-PIC simulations for problems used in [17, 18, 27, 30]. Preliminary simulations to develop and verify the approach were performed on the supercomputers MVS-10P (JSC RAS) and Lobachevsky (Lobachevsky University).

Afterwards, we collected performance data on the Endeavour supercomputer (Intel Corporation) using 384 cores of the Intel Second Generation Xeon Scalable CPU of the Cascade Lake family. Building and profiling was performed with Intel Parallel Studio XE tools. We studied QED cascade development in a circularly polarized standing wave. A slab of low density electron-positron plasma was irradiated by two counterpropagating laser beams with an amplitude of 1800 in relativistic units. A detailed description of the problem can be found in [30].

By using the improved probability estimates and avoiding over-refinement of the sub-stepping, we achieved a speedup by a factor of 8 compared to the baseline version. In addition to the qualitative analysis of the results, we measured the relative difference between the simulations using temporal evolution of two quantities: the maximal and the total macroparticle weight. Figure 2 demonstrates that these relative differences are kept under 4% for a simulation with the maximum decay probability parameter $P_{\max} = 0.01$.

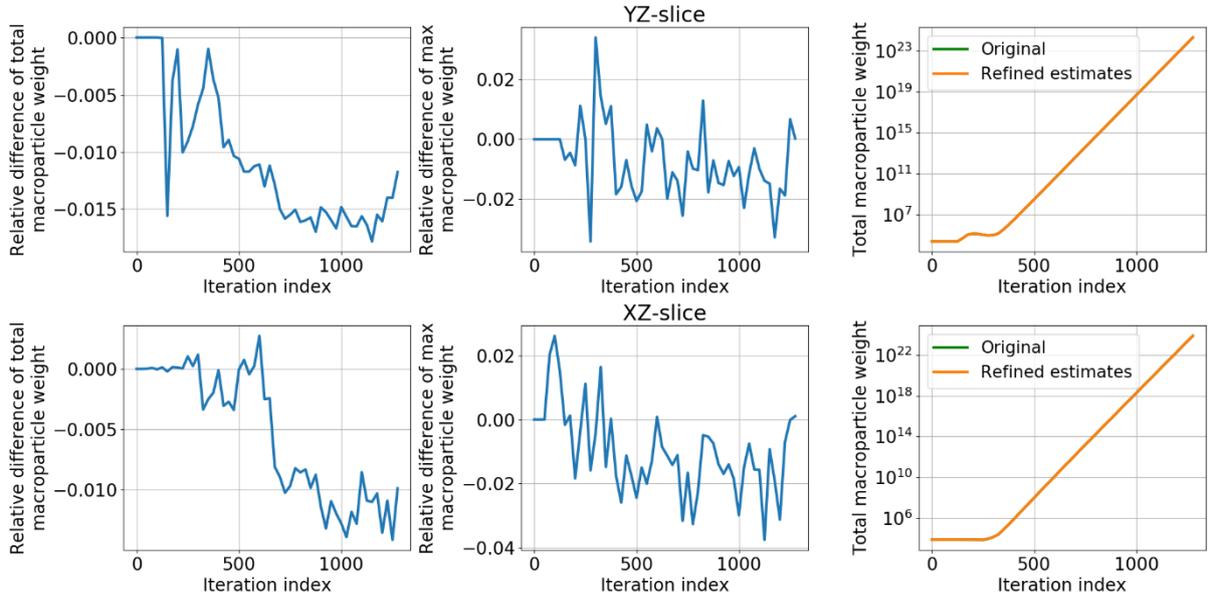

Figure 2. Relative difference between the original and improved probability estimates with $P_{\max} = 0.01$. Top row: YZ slice ($x = 0$) of the simulation domain, bottom row: XZ slice. The total macroparticle weight for the original and refined estimates on the last column is be visually overlapping due to values being very close.

Notably, the new estimates are sensitive to the $P_{\max}$ value. They were derived for values around 0.01, which we consider a reasonable choice. However, when applying it for other values of $P_{\max}$, we observed larger relative discrepancies, up to 10%.

*3.2. High-performance implementation of the synchrotron functions*

The optimization presented in the previous section allows reducing the number of time sub-steps. However, the algorithms used on each sub-step are the same, and a significant part of the computational time is spent on computing the probability formulas involving synchrotron functions.

Importantly, synchrotron function implementations are only present in few mathematical libraries. Initially, we employed the implementation for the GNU Scientific Library (GSL), with synchrotron function calculations taking 21% of the total CPU time.

Since that implementation is open source, we analyzed the source code and concluded it performs an approximation via calculating the Chebyshev value using Chebyshev polynomials up to degree 22. To improve the performance, we added pre-computing of the Chebyshev polynomial coefficients, and implemented polynomial evaluation using the Horner's method to employ fused multiply-add (FMA) instructions. It yielded a speed up of over 1.5x compared to the original GSL implementation.

The GSL implementation reproduces synchrotron function values with very high accuracy: the relative error is under $10^{-14}$. In QED-PIC simulation, the synchrotron functions are only used to compute probabilities to be used for sampling in line 6 of Algorithms 1 and 2. Thus, in our case the accuracy requirements for synchrotron functions are not nearly as strict and having a $10^{-8}$ relative error is sufficient. However, the GSL implementation is not parametrized for accuracy, and so does not offer such an option. Thus, we developed our own approximation to compute synchrotron functions, that is parametrized by the relative error bound. The approach is presented below.

*3.2.1. Method*

The synchrotron functions are defined using the modified Bessel functions of the first kind $I_\alpha(x)$ and the second kind $K_\alpha(x)$. The Bessel functions are defined as follows:

$$I_\alpha(x) = \sum_{n=0}^{\infty} \frac{1}{n!\,\Gamma(n+\alpha+1)} \left(\frac{x}{2}\right)^{2n+\alpha}$$

$$K_\alpha(x) = \frac{\pi}{2} \cdot \frac{I_{-\alpha}(x) - I_\alpha(x)}{\sin(\alpha\pi)}.$$

The first synchrotron function is $F(x) = x \int_x^{+\infty} K_{5/3}(t)\,dt$, $x \geq 0$.

The second synchrotron functions is $G(x) = xK_{2/3}(x) = \frac{\pi x}{2} \cdot \frac{I_{-2/3}(x) - I_{2/3}(x)}{\sin\left(\frac{2\pi}{3}\right)}$, $x \geq 0$.

The approach presented in this paper is applicable to both synchrotron functions, we will describe it for the second function $G(x)$.

Since the behavior of synchrotron functions is rather complicated, for approximation it is common to subdivide the range of the argument and provide a separate approximation for each interval [28]. To study asymptotic behavior of $G(x)$ for small x it is useful to consider an auxiliary function $x^{-1/3}G(x)$. Its plot for small x has the convex and concave parts. So for approximation we subdivide the range and consider $0 < x < 0.001$ and $x \geq 0.001$ separately. For relatively large values of x a separate treatment is also needed. Thus, we have three intervals of x with a separate approximation each:

- Small value of the argument: $x < 0.001$.
- Medium values of the argument: $0.001 \leq x < 4$.
- Large values of the argument: $x \geq 4$.

For small values of $x$, we use a partial sum of the series from the function definition. It can be written as:

$$S(x) = \frac{\pi x}{2\sin\left(\frac{2\pi}{3}\right)} \sum_{n=0}^{k} \frac{1}{n!} \left(\frac{x}{2}\right)^{2n} \left( \frac{\left(\frac{x}{2}\right)^{-2/3}}{\Gamma\left(n+\frac{1}{3}\right)} - \frac{\left(\frac{x}{2}\right)^{2/3}}{\Gamma\left(n+\frac{5}{3}\right)} \right),$$

where $k$ is a parameter controlling the approximation error.

For medium argument values, on the contrary, we group the $I_{-2/3}(x)$ and $I_{2/3}(x)$ terms separately. The former is transformed to obtain a series with integer degrees of $x$:

$$\frac{\pi x I_{-2/3}(x)}{2\sin\left(\frac{2\pi}{3}\right)} = x^{1/3} \sum_{n=0}^{\infty} \frac{\pi}{n!\,\Gamma\left(n+\frac{1}{3}\right)\sin\left(\frac{2\pi}{3}\right) 2^{1/3}} \left(\frac{x}{2}\right)^{2n}.$$

We introduce an auxiliary function

$$M(x) = x^{-1/3} \cdot \frac{\pi x I_{-2/3}(x)}{2 \sin\left(\frac{2\pi}{3}\right)} = \sum_{n=0}^{\infty} \frac{\pi}{n!\, \Gamma\left(n + \frac{1}{3}\right) \sin\left(\frac{2\pi}{3}\right) 2^{1/3}} \left(\frac{x}{2}\right)^{2n}.$$

For this function $M(x)$ we construct an approximating polynomial $\tilde{M}(x)$ with even degree terms of $x$, as only those are present in the original function. $\tilde{M}(x)$ can be constructed with the minimax polynomial approximation method, for example using the Remez algorithm [29].

For the $I_{2/3}(x)$ term we apply a similar approach, and use another auxiliary function

$$P(x) = x^{-5/3} \frac{\pi x I_{2/3}(x)}{2 \sin\left(\frac{2\pi}{3}\right)} = \sum_{n=0}^{\infty} \frac{\pi}{n!\, \Gamma\left(n + \frac{5}{3}\right) \sin\left(\frac{2\pi}{3}\right) 2^{5/3}} \left(\frac{x}{2}\right)^{2n}.$$

For this function $P(x)$ we again construct an approximating even-degree polynomial $\tilde{P}(x)$.

At last, for large values of the argument, the asymptotic behavior is described with the form

$$L(x) = G(x)\, x^{-1/2} e^x.$$

Function $L(x)$ behaves similarly to $\frac{1}{x}$, so for approximation we use a minimax polynomial $\tilde{L}(x)$ of variable $\frac{1}{x}$, which allows more accurate approximation.

The final form of our approximation for the second synchrotron function is

$$\tilde{G}(x) = \begin{cases} S(x), & x < 0{,}001 \\ x^{1/3}\tilde{M}(x) - x^{5/3}\tilde{P}(x), & 0{,}001 \leq x < 4 \\ x^{1/2} e^{-x} \tilde{L}(x), & x \geq 4. \end{cases}$$

*3.2.2. Implementation and Numerical results*

We calculated the degrees of approximation polynomials $\tilde{M}(x)$, $\tilde{P}(x)$, $\tilde{L}(x)$ to get the overall relative approximation error of $G(x)$ to be below $10^{-8}$ and below $10^{-14}$, that roughly correspond to the single- and double-precision floating-point accuracy. The results are presented in Table 1. Figure 3 shows the relative error plots for the $10^{-8}$ upper bound.

Then we analyze the performance of our implementation relative to the optimized GSL implementation. When using the $10^{-14}$ upper bound for the relative error, so that the accuracy is comparable, our implementation turned out to be 10% faster. For the $10^{-8}$ relative error, our implementation was 18% faster than the optimized GSL implementation. Such relaxation of accuracy requirements is enabled by QED-PIC specifics, as the synchrotron function values are only used to compute probabilities for later sampling in Algorithms 1 and 2. They do not otherwise affect the particle or field data of the simulation.

Finally, Fig. 4 presents the progress of step-by-step performance optimization. We started from the baseline version and continuously improved performance by means of estimate refinement, high-performance implementation of the synchrotron functions, developing custom approximation for synchrotron functions, and finally reducing the accuracy requirements to the sufficient level. Our improvements concerned only the particle push stage, which includes simulating the QED part and took the majority of simulation time. In terms of the total run time of the whole simulation, the overall speedup of the final version compared to the baseline implementation is 13.7 times.

We consider this speedup substantial for practical use of QED-PIC simulations. An important computational distinction of QED PIC simulations compared to traditional PIC ones is a limited potential for scaling with the number of computational nodes. Both types of PIC simulations traditionally employ spatial domain decomposition with distribution of particle and grid data, however QED-PIC often features extremely imbalanced distributions of macroparticles in the simulation area [14, 17, 18, 27, 30]. Thus, after a certain point it is challenging to proportionally reduce QED-PIC simulation time by using more computational resources. Optimization techniques presented in this paper, however, are applied internally for the most time-consuming stage and are orthogonal to techniques of domain decomposition and load balancing.

Table 1. The number of monomials to achieve given relative approximation error of the second synchrotron function.

| Relative error of $\tilde{G}(x)$ | $\tilde{M}(x)$ | $\tilde{P}(x)$ | $\tilde{L}(x)$ |
|---|---|---|---|
| $10^{-14}$ | 11 | 11 | 12 |
| $10^{-8}$ | 8 | 8 | 5 |

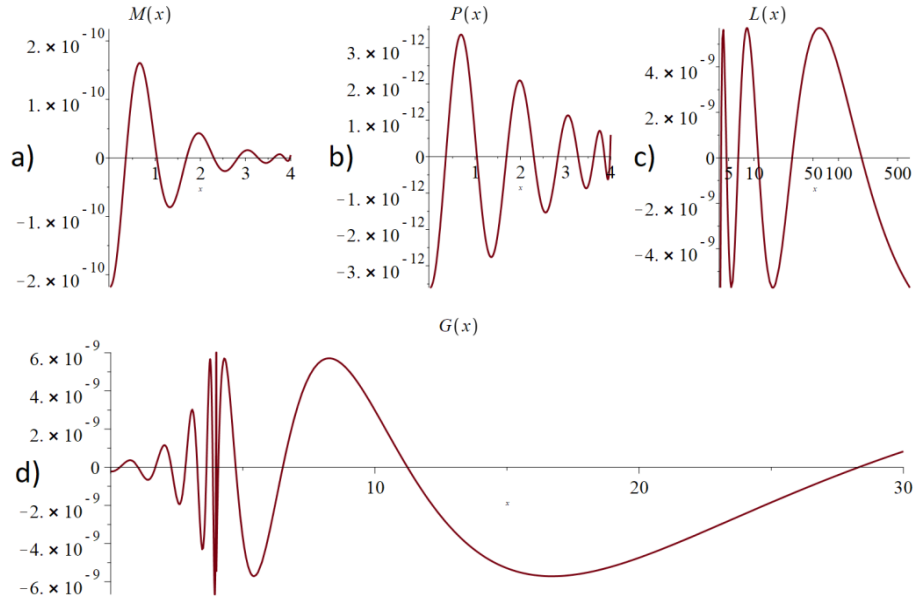

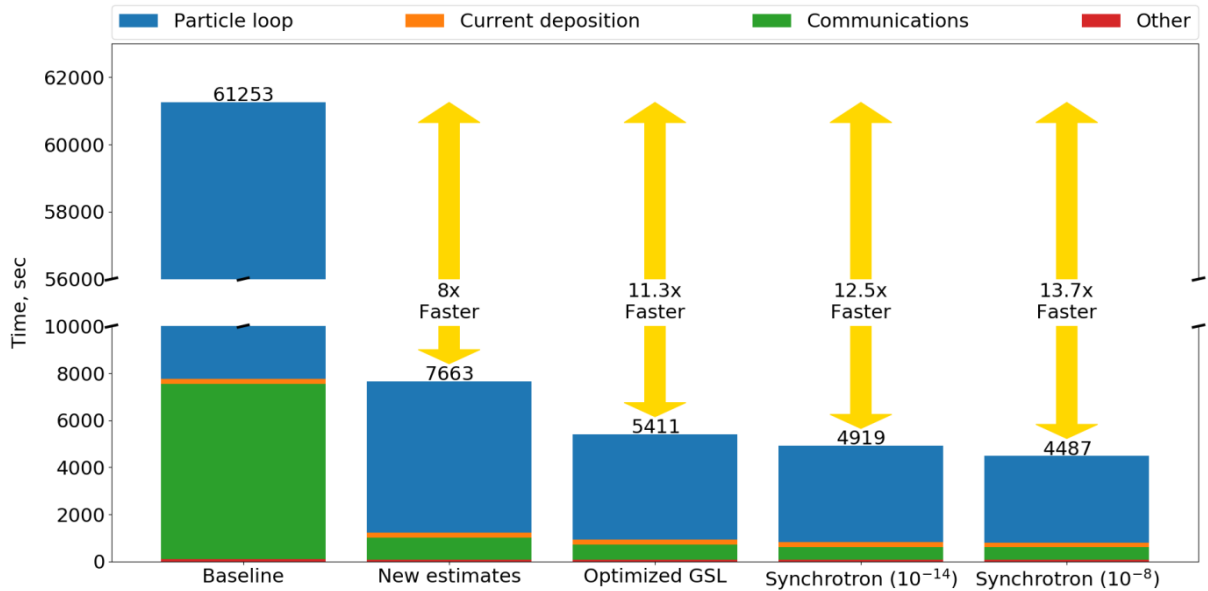

Figure 3. Relative approximation errors for the second synchrotron function $G(x)$ with the $10^{-8}$ upper bound and the auxiliary functions: a) relative error of approximating $M(x)$ with $\tilde{M}(x)$; b) relative error of approximating $P(x)$ with $\tilde{P}(x)$; c) relative error of approximating $L(x)$ with $\tilde{L}(x)$; d) relative error of approximating $G(x)$ with $\tilde{G}(x)$ for medium values of the argument.

Figure 4. Progress of step-by-step performance optimization. Sub-stepping and handling of QED events (algorithms 1, 2) is part of the particle loop stage.

## 4. Conclusions

This paper describes our approach to speed up QED-PIC simulations by a few times. Our advancements concern two directions. First, we present more accurate probability estimates to avoid the over-refinement of time sub-stepping when simulating QED cascades. Second, we constructed a new approximation of synchrotron functions and implemented it. Computational experiments with state-of-the-art QED simulations showed a substantial speedup by a factor of 13.7 in terms of the total simulation time. We have observed comparable speedups for other problems simulated with QED-PIC. The implementation is integrated into the PICADOR and Hi-Chi codes, the latter of which is distributed publicly (https://github.com/hi-chi/pyHiChi).

## 5. Acknowledgments
The work was funded by Russian Foundation for Basic Research and the government of the Nizhny Novgorod region of the Russian Federation, grant No. 18-47-520001 and the Ministry of Science and Higher Education of the Russian Federation, project no. 0729-2020-0055.